\documentclass[twocolumn]{aastex62}
\pdfoutput=1 
\usepackage{amsmath,amstext}
\usepackage[T1]{fontenc}
\usepackage[figure,figure*]{hypcap}
\graphicspath{{./}{figures/}}

\received{?}
\revised{?}
\accepted{?}
\submitjournal{AAS}

\shorttitle{Prevalence of (Near)-Resonant Orbits}
\shortauthors{Dai et al.}

\begin{document}

\title{The Prevalence of Resonance Among Young, Close-in Planets}

\correspondingauthor{Fei Dai}
\email{fdai@hawaii.edu}

\author[0000-0002-8958-0683]{Fei Dai}
\affiliation{Institute for Astronomy, University of Hawai`i, 2680 Woodlawn Drive, Honolulu, HI 96822, USA}
\affiliation{Division of Geological and Planetary Sciences,
1200 E California Blvd, Pasadena, CA, 91125, USA}
\affiliation{Department of Astronomy, California Institute of Technology, Pasadena, CA 91125, USA}

\author[0000-0003-3868-3663]{Max Goldberg}
\affiliation{Laboratoire Lagrange, Université Côte d'Azur, CNRS, Observatoire de la Côte d'Azur, Boulevard de l'Observatoire, 06304 Nice Cedex 4, France}
\affiliation{Department of Astronomy, California Institute of Technology, Pasadena, CA 91125, USA}

\author[0000-0002-7094-7908]{Konstantin Batygin}
\affiliation{Division of Geological and Planetary Sciences,
1200 E California Blvd, Pasadena, CA, 91125, USA}

\author[0000-0002-4284-8638]{Jennifer van Saders}
\affiliation{Institute for Astronomy, University of Hawai`i, 2680 Woodlawn Drive, Honolulu, HI 96822, USA}

\author[0000-0002-6246-2310]{Eugene Chiang}
\affiliation{Astronomy Department, Theoretical Astrophysics Center, and Center for Integrative Planetary Science, University of California
Berkeley, Berkeley, CA 94720, USA}
\affiliation{Department of Earth and Planetary Science, University of California, Berkeley, CA 94720, USA}

\author[0000-0003-0690-1056]{Nick Choksi}
\affiliation{Astronomy Department, Theoretical Astrophysics Center, and Center for Integrative Planetary Science, University of California
Berkeley, Berkeley, CA 94720, USA}

\author[0000-0001-9222-4367]{Rixin Li}
\altaffiliation{51 Pegasi b Fellow}
\affiliation{Astronomy Department, Theoretical Astrophysics Center, and Center for Integrative Planetary Science, University of California
Berkeley, Berkeley, CA 94720, USA}

\author[0000-0003-0967-2893]{Erik A. Petigura}
\affiliation{Department of Physics \& Astronomy, University of California Los Angeles, Los Angeles, CA 90095, USA}

\author[0000-0003-0742-1660]{Gregory J. Gilbert}
\affiliation{Department of Physics \& Astronomy, University of California Los Angeles, Los Angeles, CA 90095, USA}

\author[0000-0003-3130-2282]{Sarah C. Millholland}
\affiliation{MIT Kavli Institute for Astrophysics and Space Research, Massachusetts Institute of Technology, Cambridge, MA 02139, USA}

\author[0000-0002-3401-7463]{Yuan-Zhe Dai}
\affiliation{School of Astronomy and Space Science, Nanjing University, 163 Xianlin Avenue, Nanjing, 210023, People's Republic of China}
\affiliation{Key Laboratory of Modern Astronomy and Astrophysics, Ministry of Education, Nanjing, 210023, People's Republic of China}

\author[0000-0002-0514-5538]{Luke Bouma}
\affiliation{Department of Astronomy, California Institute of Technology, Pasadena, CA 91125, USA}

\author[0000-0002-3725-3058]{Lauren M. Weiss}
\affiliation{Department of Physics and Astronomy, University of Notre Dame, Notre Dame, IN 46556, USA}

\author[0000-0002-4265-047X]{Joshua N. Winn}
\affiliation{Department of Astrophysical Sciences, Princeton University, 4 Ivy Lane, Princeton, NJ 08544, USA}



\begin{abstract}
\noindent

Multiple planets undergoing disk migration may be captured into a chain of mean-motion resonances with the innermost planet parked near the disk's inner edge. Subsequent dynamical evolution may disrupt these resonances, leading to the non-resonant configurations typically observed among {\it Kepler} planets that are Gyrs old. In this scenario, resonant configurations are expected to be more common in younger systems.  This prediction can now be tested, thanks to recent discoveries of young planets, particularly those in stellar clusters, by NASA's {\it TESS} mission. We divided the known planetary systems into three age groups: young ($<$100-Myr-old), adolescent (0.1-1-Gyr-old), and mature ($>1$-Gyr-old). The fraction of neighboring planet pairs having period ratios within a few percent of a first-order commensurability (e.g.~4:3, 3:2, or 2:1) is 70$\pm$15\% for young pairs, 24$\pm$8\% for adolescent pairs, and 15$\pm$2\% for mature pairs. The fraction of systems with at least one nearly commensurable pair (either first or second-order) 
is 86$\pm13$\% among young systems, 38$\pm12$\% for adolescent systems, and 23$\pm3$\% for mature systems. First-order commensurabilities prevail across all age groups, with an admixture of second-order commensurabilities. Commensurabilities are more common in systems with high planet multiplicity and low mutual inclinations. Observed period ratios often deviate from perfect commensurability by $\sim$1\% even among young planets, too large to be explained by resonant repulsion with equilibrium eccentricity tides. We also find that super-Earths in the radius gap ($1.5-1.9R_\oplus$) are less likely to be near-resonant (11.9$\pm2.0\%$) compared to Earth-sized planets ($R_p<1R_\oplus$; 25.3$\pm4.4\%$) or mini-Neptunes ($1.9R_\oplus \leq R_p<2.5R_\oplus$; 14.4$\pm1.8\%$). 
\end{abstract}


\section{Introduction}

\begin{deluxetable*}{|l|l|l|l|}
\caption{Terminology used in this paper}
\label{tab:terminology}
\tablehead{
\colhead{Term} & \colhead{Definition} }
\startdata
1st-Order Resonant Pair & Two planets near a 6:5, 5:4, 4:3, 3:2, or 2:1 period commensurability with $-0.015 \leq \Delta \leq +0.03$
\\
\hline
2nd-Order Resonant Pair & Two planets near a 3:1, 5:3, 7:5, or 9:7 period commensurability with $-0.015 \leq \Delta \leq +0.015$ 
\\
\hline
Singly-Resonant System & System with only one resonant pair (1st-order or 2nd-order) as defined above \\
\hline
Multi-Resonant System & System with two or more resonant pairs (1st-order or 2nd-order) as defined above \\
\hline
Resonant Chain & A multi-resonant system with at least two contiguous resonant pairs (e.g.~three planets with period ratios of 1:2:6)\\
\enddata
\tablecomments{The bounds on $\Delta$ used to define resonance are motivated by observations (see Fig.~\ref{fig:delta} and discussion in Section 3). In and of itself, $\Delta$ cannot formally determine whether a resonant argument is librating or circulating.
Nevertheless, most 1st-order pairs with $\Delta \lesssim 0.6\%$ have been argued to be librating \citep{Goldberg2023}, and most 1st-order pairs with $\Delta \sim +1$-3\% (comprising the majority of our ``resonant'' sample) are thought to have been originally captured into libration, and to have been perturbed since into large-amplitude libration or circulation \citep{Lithwick_repulsion,Batygin_repulsion,Choksi2023,Chatterjee_2015,Wu2024}.
}
\end{deluxetable*}

As multiple planets form and gravitationally interact with a protoplanetary disk, they are expected to migrate,  usually towards the central star \citep{Goldreich1979,Ward,Lin1986,McNeil,Terquem_2007,Nelson2018}. A pair of neighboring planets may be captured into a mean-motion resonance (MMR) if they have low eccentricities \citep{Borderies}, and if their relative migration rate is adiabatic (occurring on a timescale longer than the resonant libration period) and convergent (orbital period ratio decreases over time).

The gradient in the surface density of the disk near its inner edge may be responsible for halting inward migration \citep{Masset_2006}. As the innermost planet waits at the disk edge, longer-period planets can catch up, converting otherwise divergent encounters into convergent ones. Population synthesis models that incorporate Type I migration routinely produce chains of close-in planets in MMR \citep[e.g.][]{Ormel2017,Ogihara2018,Dai1136,Wong2024}. This process of resonant capture
is so effective that the majority of ‘Kepler-like’ close-in, multi-planet systems could have initially formed as
resonant chains \citep{Izidoro,Emsenhuber2021}. 

On the other hand, only a relatively small fraction of mature {\it Kepler} planets \citep[median age 4.6-Gyr-old,][]{Berger_age}  are currently near first-order MMRs \citep[14.8$^{+0.5}_{-0.7}\%$;][]{Huang2023}. Once the gaseous disk dissipates, the orbital eccentricities and mutual inclinations of planets are no longer damped by interactions with the disk \citep[see e.g.,][]{Papaloizou2000}. Various dynamical processes start to disrupt the initially resonant configuration produced by disk migration. Proposed processes include overstable libration \citep[from post-disk eccentricity damping, e.g. from tides][]{Goldreich2014,Xu0217}, 
planetesimal scattering \citep{Chatterjee_2015,Wu2024},  divergent migration due to disk dispersal \citep{Hansen2024,Liu2022},
secular chaos \citep{Petrovich}, and orbital instability \citep{PuWU,Izidoro,Goldberg2022}. 

\citet{Goldberg2022} suggested that orbital instability may be responsible for the observed intra-system uniformity of planet sizes and regular orbital spacing which is widely known as the  `peas-in-a-pod' pattern \citep{Weiss_peas,Millholland_peas,Wang_uniform,Goyal2023}. See also the simulations by \citet{MacDonald2020}, \citet{Izidoro2022}, \citet{Weiss2023}, and \citet{Lammers_uniformity}. In the same vein, the outer Solar System could have been initially resonant \citep{Morbidelli2007,Thommes2008,Batygin2010,Nesvorny2012} before being disrupted by planetesimal scattering and ensuing instability, as proposed by the Nice model \citep[][]{Gomes,Tsiganis,Morbidelli2005}. 

Recently, \citet{Hamer2024} showed using {\it Gaia} data \citep{Gaia} that the space velocity dispersion of host stars with near-resonant planets, especially those near second-order resonances, is lower than that of the field stars. A lower kinematic dispersion suggests youth, but kinematic ages are only accurate on Gyr timescales \citep[e.g.][]{Aumer,Yu2018,Ting2019}. Alternative methods are needed to reveal individual planetary systems younger than 100s of Myr. Thanks to {\it TESS}'s nearly full-sky coverage \citep{Ricker}, dozens of young planetary systems have been discovered, many of which are cluster members with precise age constraints of tens to hundreds of Myr. We can now test the paradigm of close-in planet formation described above: are young planetary systems more likely to be in near-resonant configurations? 

Previous works have mostly focused on first-order resonances \citep{Fabrycky2014,Izidoro,Huang2023}. In this work, we include second-order resonances as well. Theoretically, higher-order resonances are weaker at low eccentricities and should be rarer. Nonetheless, \citet{Xu0217} demonstrated that second-order resonances are possible if migration is slow enough and the planets have similar masses. Second-order resonances have been suggested for Kepler-29 \citep[9:7, ][]{Fabrycky_Kepler29,Migaszewski_2017,Cui}, and have been observed in TOI-1136 \citep[7:5, ][]{Dai1136}. Second and higher-order resonances have been observed in the Solar System as well, particularly among the satellites of gas giants \citep{Murray}.

We assemble the latest list of confirmed, young planetary systems in Section 2. We identify the near-resonant planets in Section 3. We examine the prevalence of resonances across ages and system architectures in Section 4. We discuss the potential caveats of this work and offer some ideas for future work in Section 5. The paper is summarized in Section 6.

\section{A Sample of Young Multi-planet Systems}\label{sec:sample}
The foundation of our investigation is a sample of young, multi-planet systems with precisely measured ages. We queried the NASA Exoplanet Archive \footnote{\url{https://exoplanetarchive.ipac.caltech.edu}} on Jan 26 2024 for the latest sample of confirmed exoplanetary systems. We restricted our attention to transiting planets since their orbital periods are known with much higher precision than for non-transiting planets. We divided the sample into three broad-brush age groups: 1) young $<$100-Myr, 2) adolescent 100-1000-Myr old; 3) mature $>$1-Gyr. Precise stellar ages are usually difficult to obtain. For a comprehensive discussion of the applicability and limitations of commonly used age indicators, readers are referred to the literature on isochronal fitting \citep[e.g.][]{Berger_age}, chromospheric activity \citep[e.g.][]{Mamajek}, gyrochronology \citep[e.g.][]{vanSaders2016,Bouma_gyro}, lithium abundances \citep[e.g.][]{Berger_li,Bouma_li}, kinematics \cite[e.g.][]{Hamer2020,Chen_past}, and astereoseismology \citep[][]{Metcalfe}. These age indicators often come with significant uncertainties, sometimes amounting to half an order of magnitude. Cluster membership likely provides the most precise and well calibrated age constraints. However, cluster membership becomes increasingly difficult to establish for ages older than 1 Gyr, when the clusters become more diffuse \citep[e.g.][]{2010ARA&A..48..431P,2019ARA&A..57..227K,2020A&A...640A...1C}. 

Although we started with the ages reported in the literature, we closely examined the age constraints on a case-by-case basis. The youngest group ($<$ 100 Myr) in our sample plays a pivotal role. We have 7 multi-planet systems whose ages primarily come from their cluster membership. TOI-6109 \citep[Dattilo et al. in prep]{YZDai,Vach} is a member of the Alpha Persei cluster / Theia 133 association \citep[$\sim$55-Myr-old, see also $\sim80$-Myr-old estimate by][]{Boyle}. TIC 434398831 \citep{Vach} belongs to Theia 116 ($\sim$48-Myr-old). HD 109833 \citep{Wood} is a member of the Lower Centaurus Crux Association (10 to 16-Myr-old). V1298 Tau \citep{David2019} belongs to the Taurus-Auriga association ($\sim23$-Myr-old). We used the updated transit ephemeris of V1298 Tau from Livingston et al. (in prep.). AU Mic \citep{Plavchan} is a pre-main-sequence star with a debris disk in the $\sim$20-Myr-old $\beta$~Pic moving group. TOI-1227 \citep{Mann1227} is a member of the  Lower Centaurus Crux/Musca Group ($\sim$11-Myr-old). It only has one detected transiting exoplanet. However, TTVs analyzed by \citet{Almenara1227} revealed the presence of another planet likely in a 3:2 MMR with the known planet. Although TOI-942 \citep{Zhou942,Carleo} is not a cluster member, various age indicators and different groups of researchers agree that the age is $\sim$50-Myr. 

For the adolescent systems of 100 Myr - 1 Gyr, we have a sample of 16 multi-planet systems (Fig. \ref{fig:1gyr}). K2-136 \citep{Mann136,Ciardi136} is a member of the Hyades ($\sim$650-Myr-old). HD 63433 \citep{Mann63433,Capistrant} is a member of Ursa Major moving group ($\sim$400-Myr-old). TOI-451 \citep{Newton451} is a member of the Pisces-Eridanus Stream ($\sim$120-Myr-old). K2-264 is a member of Praesepe \citep{Rizzuto264,Livingston264}. TOI-2076 \citep{Hedges} was shown to be comoving with TOI-1807; jointly their ages were constrained to be $\sim$200-Myr-old. 

Fortunately, gyrochronology is well-calibrated in the 100 Myr - 1 Gyr age range. This is because stars have had significant time to spin down, and many open clusters are available for calibration \citep{Bouma_gyro}. We independently constrained the ages of stars in this age group with our own gyrochronological analysis. 
We used the braking model of \citet{vanSaders2016} and the pre-generated grid from \citet{Saunders}. Using the sampling strategy described by \citet{Metcalfe}, we incorporated the observed constraints on effective temperature, [Fe/H], luminosity, and current-day rotation period of the stars in our sample. Our findings confirmed that these stars are indeed younger than 1 Gyr, consistent with the literature.


For comparison, we also identified a group of mature planetary systems ($>1$-Gyr-old). 
There are 255 mature systems reported in the literature. Most of the age constraints in this age group come from isochrone fitting. \citet{Berger_age} showed that the median age of the Kepler planet hosts is 4.6 Gyr, similar to the age of the Solar System. Finally, we applied the same set of analyses to all confirmed multi-planet systems, many of which do not have reported ages. There are 734 of these (see Tab. \ref{tab:mmr}). 

To check that any trend we report in this work is not driven by system parameters other than the age, we compared the distribution of stellar parameters ($T_{\rm eff}$, $M_\star$, [Fe/H]) and planetary parameters ($R_p$, $P_{\rm orb}$) between the different age groups. Planet mass $M_p$ is not available for the majority of confirmed planets. We performed both Anderson-Darling and Kolmogorov-Smirnov tests implemented in {\sc scipy} on the various stellar and planetary parameters that are available. To account for measurement uncertainties, we bootstrapped using the uncertainties reported in the literature. The only parameter that is statistically distinct (p-value $<5\%$) between the age groups is the planetary radius. Both the young ($<$100-Myr-old) and the adolescent (0.1-1 Gyr-old) planets have larger radii than the mature planets ($>$1-Gyr-old). 

\begin{figure*}
\center
\includegraphics[width = 2.\columnwidth]{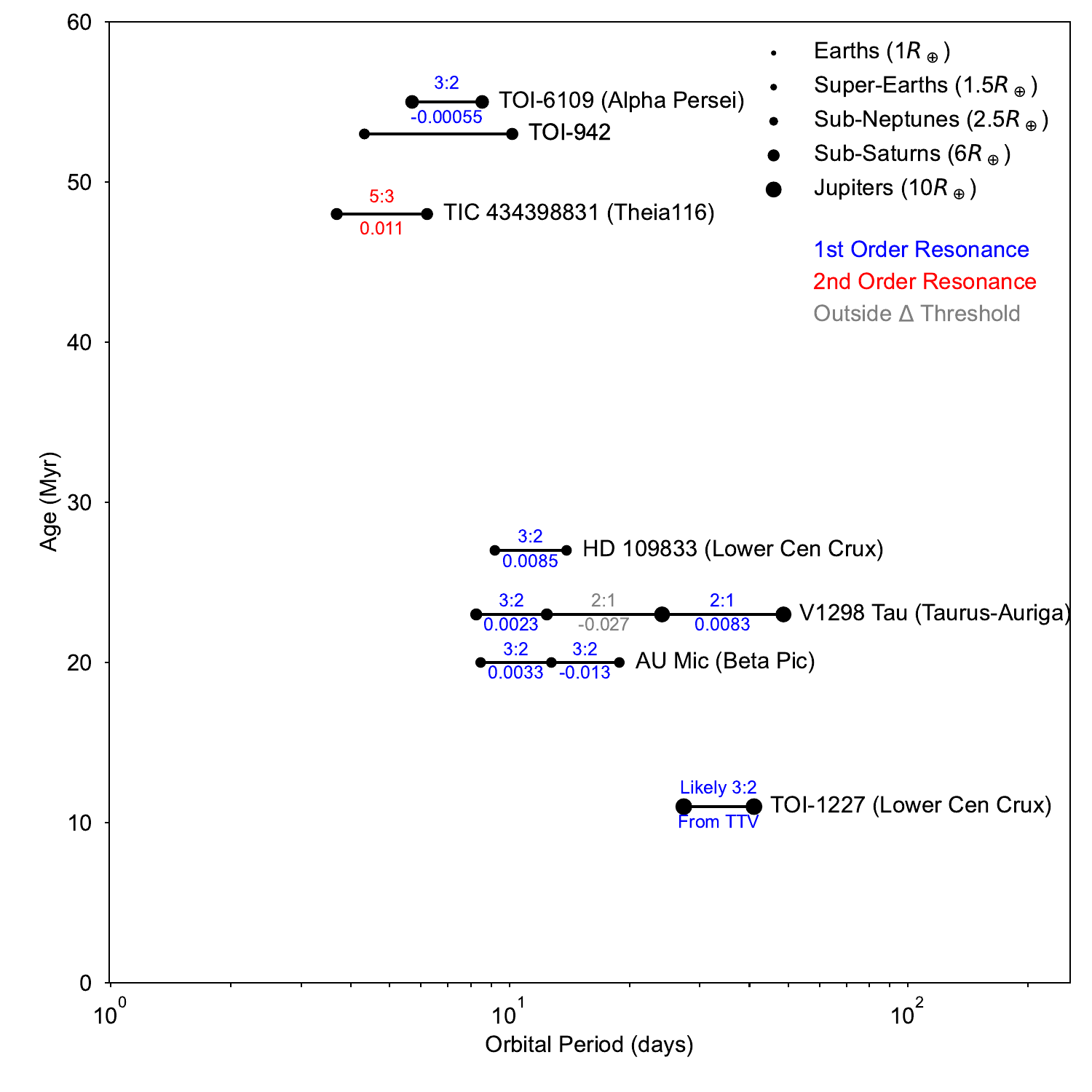}
\caption{Orbital architectures of `young' multi-planet systems ($<$ 100~Myr old). Cluster members are labeled. Symbol sizes represent planet radii as indicated. Each near-resonant pair of planets is labeled by its resonance and the fractional deviation $\Delta$ from perfect period commensurability (eqn.~\ref{eqn:delta}). Colors indicate resonance order. Planet pairs situated within several percent of a period commensurability but that fall formally outside our $\Delta$ thresholds (Tabs.~\ref{tab:terminology} and \ref{tab:mmr}) are colored light gray (and not counted in our resonance incidence rates). Near-resonant configurations appear to dominate these young planetary systems, with 8 out of 10 pairs of neighboring planets falling within our $\Delta$ thresholds for resonance, and 6 out of 7 systems exhibiting at least one resonant pair so defined.}
\label{fig:100myr}
\end{figure*}

\begin{figure*}
\center
\includegraphics[width = 2.\columnwidth]{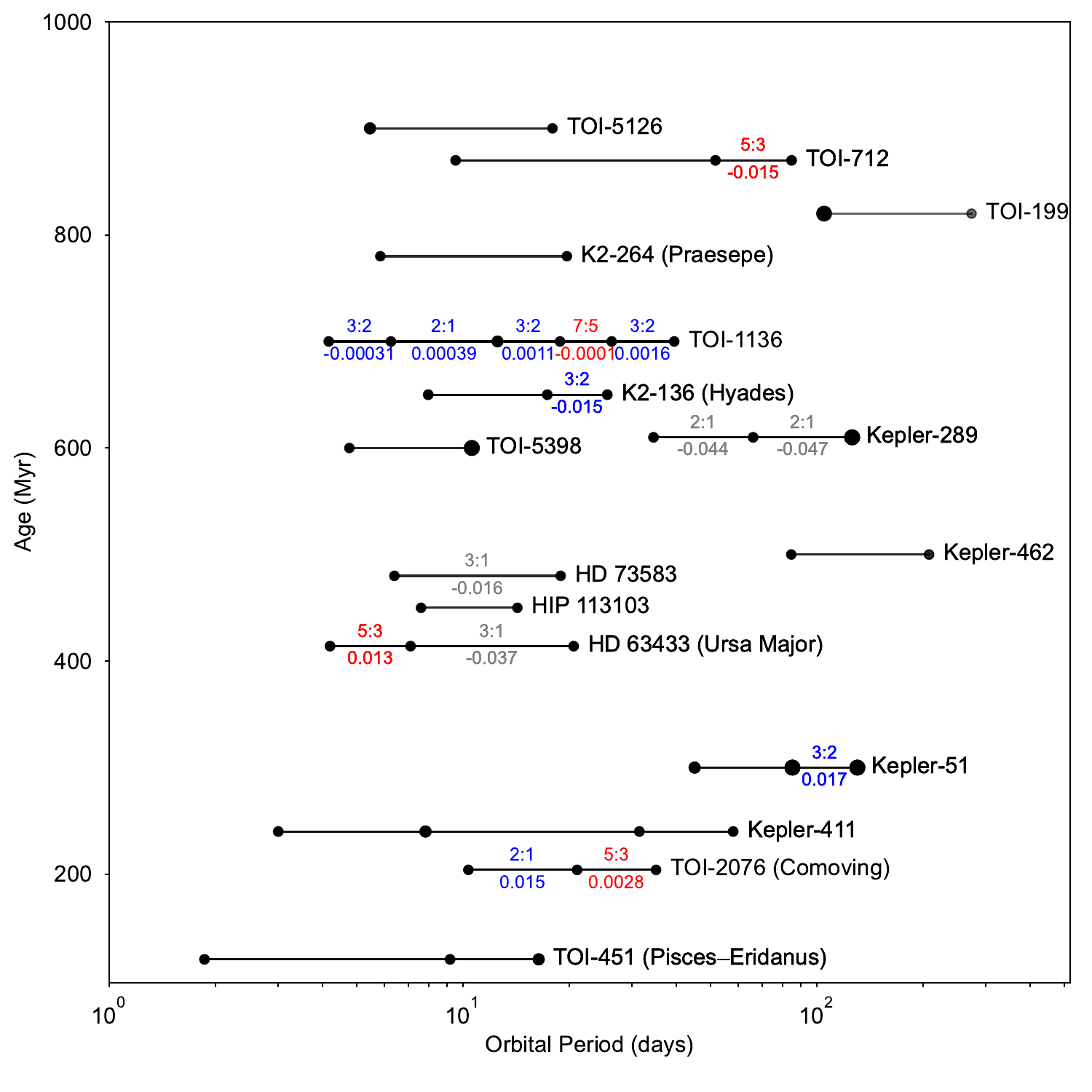}
\caption{Same as Fig.~\ref{fig:100myr} but for the `adolescent' planetary systems, defined as having ages between 100 Myr and 1 Gyr. Cluster members are labelled. 
Six out of 16 systems have at least one pair of near-resonant planets, and 11 out of 29 neighboring pairs are near-resonant (as defined by our $\Delta$ criteria; see Tabs.~\ref{tab:terminology} and \ref{tab:mmr}). The near-resonant pair of planets in TOI-712 falls right on our $\Delta$ boundary but exhibits TTV variations suggesting strongly resonant interaction (data not shown).}
\label{fig:1gyr}
\end{figure*}

Although many of the young planets appear to have large radii characteristic of mature giant planets (Fig. \ref{fig:100myr}), they are likely inflated sub-Neptune planets that are still contracting from the residual heat of their formation  \citep[see e.g.][]{Fortney}. The largest planets in the young age group are V1298 Tau b (7.9 $R_\oplus$) and e (9.5 $R_\oplus$), which likely have masses on the order of $10 M_\oplus$ according to the TTV model by Livingston et al. (in prep.). \citet{Barat} reported an upper limit of $<24~M_\oplus$ for V1298 Tau b based on a low surface gravity inferred from Hubble transmission spectroscopy of the planet's atmosphere. Additionally, the radial velocity analysis by \citet{Blunt} also suggested that the high masses reported by \citet{Mascareno} were probably incorrect.

\section{Identifying Resonant Planets}\label{sec:ident}

Having assembled a list of multi-planet systems with ages, we now attempt to identify 
resonant pairs. The most readily computed proxy for resonance is the orbital period ratio between neighboring planets, or more specifically, the fractional deviation from perfect period commensurability:
\begin{equation}\label{eqn:delta}
\Delta \equiv \frac{P_{\rm out}/P_{\rm in}}{p/q}-1   
\end{equation}
where $P_{\rm in}$ and $P_{\rm out}$ are the observed orbital periods of the inner and outer planet in a neighboring pair, and $p$ and $q$ are small positive integers that define a given MMR \citep[see e.g.][]{Papaloizou2010,Lithwick_repulsion,Batygin_repulsion}. The order of the resonance is given by $p-q$, where $p-q$ = 1, 2, 3 corresponds
to a 1st, 2nd, and 3rd-order resonance, respectively.

Small values of $\Delta$ suggest---but do not strictly imply---bodies ``librating'' in resonance, such that their resonant ``arguments'' (linear combinations of mean and periapse longitudes) oscillate with finite amplitude about certain ``libration centers'' or fixed points (see \citealt{Murray} for an introduction). In resonant libration, planetary conjunctions (close encounters) preferentially occur near specific orbital phases (e.g.~periapse or apoapse). Among present-day observed resonant chains (exhibiting multiple and contiguous resonant pairs, e.g.~1:2:4; see Table 1 for a glossary), resonant arguments have been found to librate \citep[e.g.][]{Mills2016, Agol2021, Leleu2021, MacDonald2021, Dai1136}. \citet{Goldberg2023} argued that libration is more common among planet pairs with $\Delta \lesssim 0.6\%$.

\begin{figure*}
\center
\includegraphics[width = 1.\columnwidth]{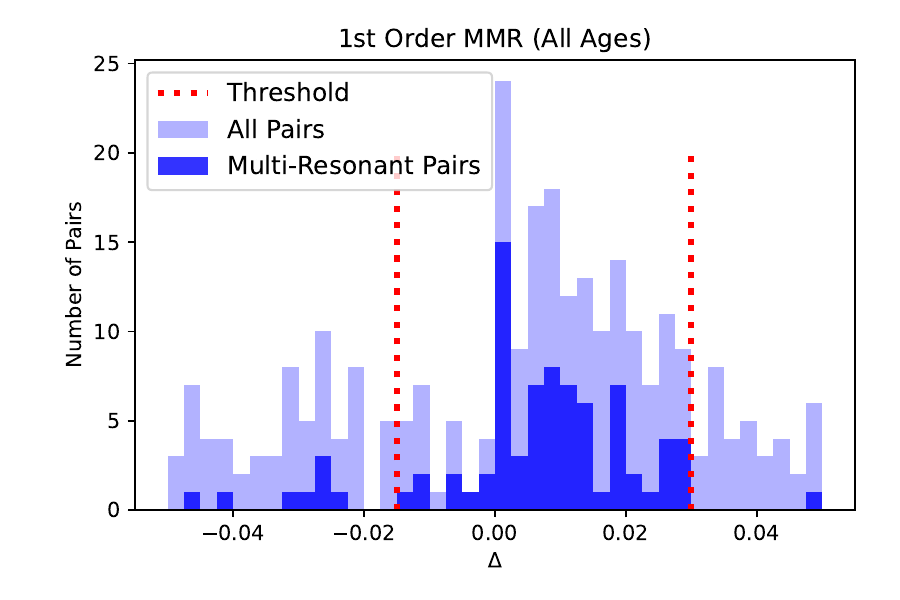}
\includegraphics[width = 1.\columnwidth]{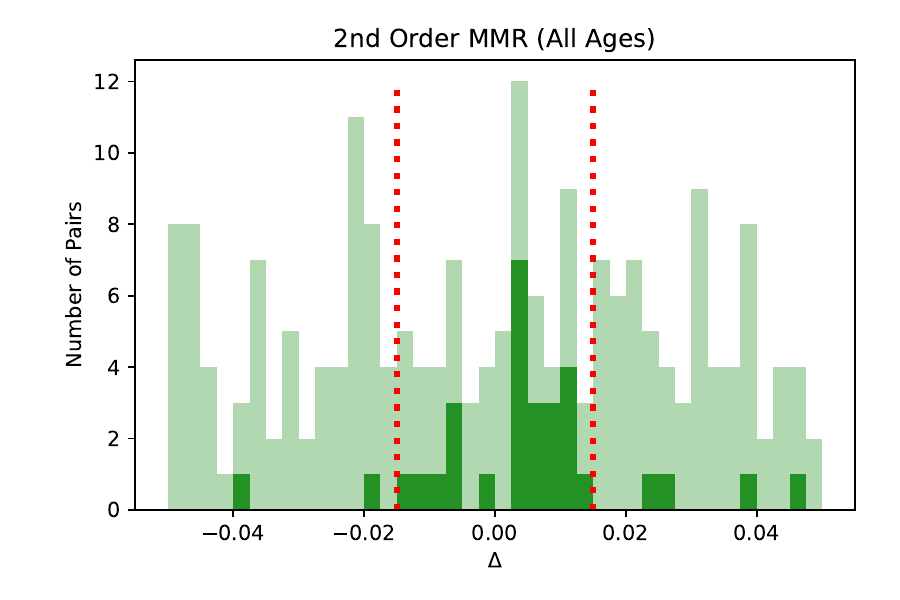}
\caption{Histograms of $\Delta$ (the deviation from perfect period commensurability; eqn. \ref{eqn:delta}) for planet pairs near 1st-order resonances (left panel including the 6:5, 5:4, 4:3, 3:2, and 2:1 MMRs) and 2nd-order resonances (right panel including the 9:7, 7:5, 5:3, and 3:1 MMRs), across all ages (known and unknown). The translucent histograms are drawn from all 734 confirmed multi-planet systems, and the dashed red lines mark our chosen bounds on $\Delta$ inside of which pairs are deemed `resonant'. We used an asymmetrical boundary $-0.015\leq \Delta \leq0.03$ for 1st-order resonances to reflect how their libration widths are intrinsically asymmetric \citep{Batygin_resonance,Choksi2020}, and a symmetrical boundary of $-0.015 \leq \Delta \leq0.015$ for 2nd-order resonances \citep{Bailey}. Note how the excess population or `peak' in the left panel is captured by our $\Delta$ bounds. Systems identified to have more than one resonant pair --- these are arguably self-validating, since resonances congregate in a system more than they would if they were distributed randomly (section \ref{sec:co-occurrence}) --- are used to generate the opaque histograms, most of which lie self-consistently within our $\Delta$ bounds (not all of the opaque points lie inside, since a given multi-resonant system can also contain non-resonant pairs).}
\label{fig:delta}
\end{figure*}

Observationally, mature {\it Kepler} planets near 1st-order resonances have been found with $\Delta$ as large as 1--3\% \citep[e.g.][]{Fabrycky2014} --- see also our Figure \ref{fig:delta}, which plots the combined  $\Delta$ distributions near the 2:1, 3:2, 4:3, 5:4, and 6:5 MMRs (left panel, translucent histogram). Fig.~\ref{fig:delta} shows the well-known excess population of pairs at positive $\Delta$ (the ``peak'' in the $\Delta$ histogram), and the accompanying deficit of pairs with negative $\Delta$ (``trough''). This peak-trough substructure has been interpreted as a signature of resonant interactions modified by energy dissipation/eccentricity damping from  dynamical friction with a surrounding gas or planetesimal disk, or tides raised on planets \citep{ Lithwick_repulsion,Batygin_repulsion,Choksi2023,Chatterjee_2015,Wu2024}. Statistical analysis of TTVs shows that mature pairs in 1st-order resonant peaks have resonant arguments that either librate with large amplitude ($\gtrsim 1$ rad), or circulate from 0 to 2$\pi$ \citep{Lithwick_ttv, Hadden2013, Goldberg2023, Choksi2023}. Such systems are theorized to have been librating with small amplitude in the past, and to have been subsequently knocked off their fixed points to be circulating, or nearly so, today. 


Guided by Fig.~\ref{fig:delta} and the above discussion, we define as ``1st-order resonant'' those pairs near a 2:1, 3:2, 4:3, 5:4, or 6:5 period commensurability with $-0.015 \leq \Delta \leq +0.03$. The boundaries in $\Delta$ are asymmetric about 0 to reflect how the libration centers of first-order resonances are intrinsically located at $\Delta > 0$ \citep{Batygin_resonance,Choksi2020}. The boundaries are wide enough to accommodate the excess population in the peak shown in Fig.~\ref{fig:delta}, thought to include pairs that experienced resonance capture in the past.

Similarly, we define as ``2nd-order resonant'' those pairs near a 3:1, 5:3, 7:5, or 9:7 period commensurability with $-0.015 \leq \Delta \leq +0.015$. The half-width of 0.015 is suggested by observations (Fig.~\ref{fig:delta}, right panel), and is held to be symmetric about 0 based on dynamical integrations by \citet{Bailey}. We do not present results on 3rd-order resonances since we could find no statistically significant excess populations in their vicinity.

As a further check on our selected bounds on $\Delta$, we also identify multi-resonant systems---those with more than one resonant pair (as defined using our $\Delta$ criterion). Multi-resonant systems are arguably `self-validating' because, as we will show in section \ref{sec:co-occurrence}, the odds of finding two resonant pairs in a given system---if resonant pairs are distributed randomly and sporadically over all systems---is low (about $0.156^2 \sim 2\%$ for two first-order pairs, and even lower for second-order pairs). The actual observational data indicate much higher incidence rates of multi-resonant pairs (section \ref{sec:co-occurrence}), suggesting they were formed by a physical mechanism (convergent migration of multiple planets within a disk). When we plot the $\Delta$ distribution of actual multi-resonant systems (Figure \ref{fig:delta}, opaque histograms), the bulk of the distribution lies self-consistently within our chosen $\Delta$ bounds.

We emphasize that our definitions of ``resonant'' (see Table 1 for a summary) are largely empirical, and based purely on $\Delta$, which in and of itself does not determine whether resonant arguments are librating or circulating. Unless indicated otherwise, we will use the terms ``resonant'', ``near-resonant'', ``commensurable'', and ``near-commensurable'' interchangeably. Note that
measurement uncertainties in $\Delta$ are probably negligible compared to our 1-$3\%$ thresholds on $|\Delta|$. Measurement errors in $\Delta$ from \textit{Kepler} are $<10^{-6}$ \citep{lissauer2023}. For \textit{K2} and \textit{TESS}, which have much shorter time baselines per target, the error in $\Delta$ is larger at $<10^{-4}$ level. Even in the most extreme case where only two transits are observed, $\Delta P/P$ could amount to 0.2\% for $\sim$10-day planet showing  $\sim 30$ minutes TTV, a typical amplitude for a near-resonant pair \citep{Hadden2017}.

\section{Results and Discussion}
Having laid down our criteria for membership in a first or second-order resonance, we now tabulate the fraction of resonant systems across our three different age bins and across orbital architectures. Table \ref{tab:mmr} summarizes the results. 

Across all ages (known and unknown), first-order resonances are more common than second-order resonances. We found that a pair of neighboring planets has a  15.6$\pm1.1\%$  chance of being close to a first-order commensurability (177 out of 1131 possible pairs satisfy our $\Delta$ criterion).  This is similar to the value of 14.8$^{+0.5}_{-0.7}\%$ reported by \citet{Huang2023}. In contrast, the fraction for second-order resonances drops to 66/1131 or 5.8$\pm0.7\%$. Among the first-order resonances, the 3:2 resonance is the most prevalent (83/1131, 7.3$\pm0.8\%$). The runner-up is 2:1 (57/1131, 5.0$\pm0.7\%$), followed by 4:3 (23/1131, 2.0$\pm0.4\%$). The 5:4 (12 out of 1131, 1.1$\pm0.3\%$) and 6:5 (2 out of 1131, 0.2$\pm0.1\%$) resonances are rarer still. Here and in the rest of the paper, we compute the fraction of resonant system assuming a binomial distribution with $P_{\rm res} = N_{\rm res}/N_{\rm tot}; \sigma_{\rm res} = \sqrt{N_{\rm res}(N_{\rm tot}-N_{\rm res})/N_{\rm tot}^{3}}$.

\begin{figure}
\center
\includegraphics[width = 1.05\columnwidth]{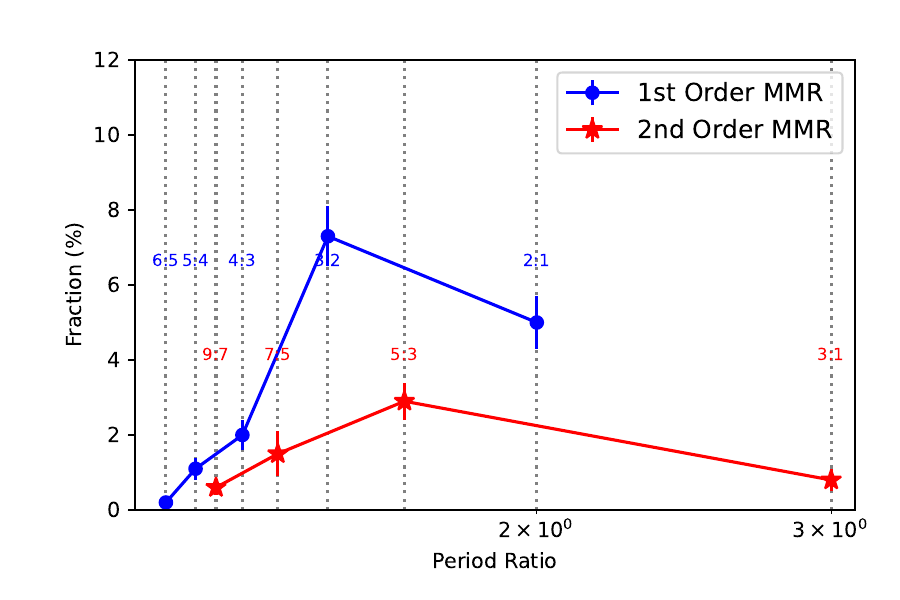}
\caption{ The fractions of observed neighboring pairs of planets across all ages near various first and second-order resonances. Resonances with smaller period ratios (4:3, 5:4, 6:5) are progressively less populated because a pair of planets must avoid all earlier resonances to reach them. Another possibility is that these resonances may start to overlap and destabilize \citep[e.g.][]{Rath}.}
\label{fig:relative_occurrence}
\end{figure}

The 2:1 resonance is weaker than the 3:2 because the 2:1 involves a larger planet-planet separation, and because the 2:1's  indirect gravitational potential (due to the reflex motion of the star) weakens its disturbing function (\citealt{MurrayClayChiang2005}). The 2:1 resonance may thus require slower migration rates (lower disk surface densities) to capture planets effectively \citep{Kajtazi,Batygin_Petit}. Additionally, overstable libration \citep[][]{Goldreich2014,Deck} may remove planets from the 2:1 resonance. The stronger 3:2 resonance may be the first-encountered resonance that robustly captures migrating planets and tends to survive for longer timescales. Higher-$p$ resonances (4:3, 5:4, 6:5) are progressively less populated because a pair of planets must avoid lower-$p$ resonances to reach them (see Fig. \ref{fig:relative_occurrence}).  Another possibility is that higher-$p$ resonances may start to overlap and destabilize \citep[e.g.][]{Hadden2018,Petit_instability，Rath}.

Among the 2nd-order resonances across all ages (known and unknown), the 5:3 resonance is the most prevalent (33/1131 or 2.9$\pm$0.5\%), with the 7:5 resonance in second place (17/1131 or 1.5$\pm$0.6\%). Only 0.8$\pm$0.3\% (9/1131) of planets are close to the relatively weak 3:1 resonances.  \citet{Bailey} examined the period ratio distribution of {\it Kepler} systems and showed that there is no statistically significant peak at the 3:1 resonance, but that there may be a peak near the 5:3 resonance. 

\subsection{Resonances Tend to Congregate in a Given System}\label{sec:co-occurrence}
Does a resonant pair appear in isolation? Or do multiple resonant pairs appear together in the same planetary system?  To answer this question, we study planetary systems with three or more detected planets across all ages.

\begin{figure*}
\center
\includegraphics[width = 2.\columnwidth]{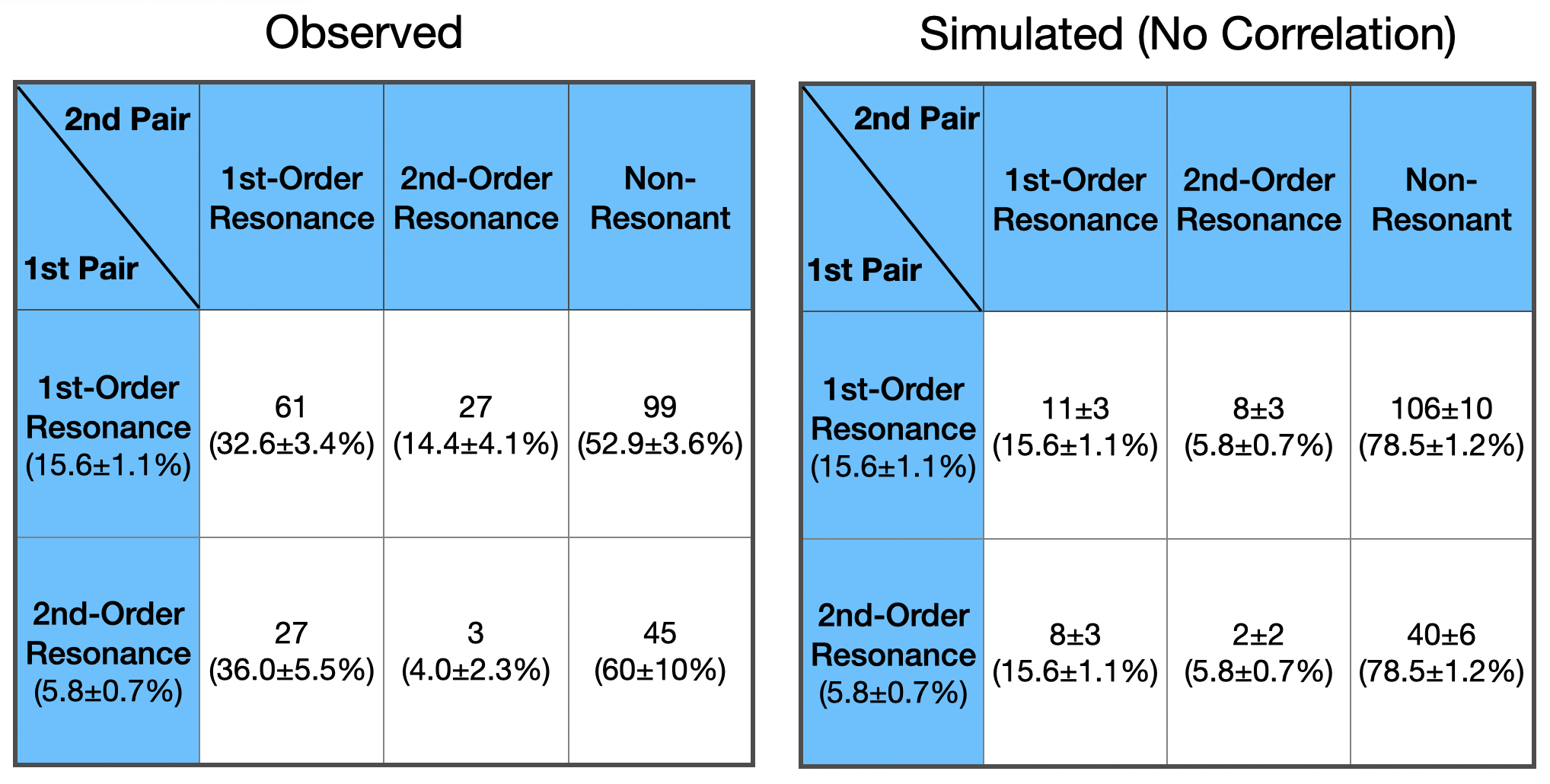}
\caption{Given a system with one resonant pair, what is the probability that this same system has a second resonant pair? We experiment using observed multi-planet systems across all ages (known and unknown). The rows indicate the given first resonant pair (either 1st-order or 2nd-order, with their raw probabilities), while the columns label the second pair. The first entry in the left panel shows that, using real observed systems, there are 61 cases where both pairs are near first-order resonances. This corresponds to a observed probability of 61/(61+27+99) = 32.6$\pm3.4$\% for finding a second 1st-order resonance given a first 1st-order resonance. This probability is twice as high as in a simulated uncorrelated sample that scrambles observed period ratios across all systems (right panel; the corresponding probability is 15.6$\pm1.1\%$, which is just the raw incidence of 1st-order resonances). Resonances among observed systems thus `congregate': where there is one resonant pair, there are better-than-random odds that there will be another in the same system.
}
\label{fig:covariance}
\end{figure*}

The matrix shown in the left panel of Fig.~\ref{fig:covariance}
tabulates the likelihoods of finding a resonant pair accompanied by another resonant pair within the same system. For instance, the first three entries in the first row show the numbers of additional resonant pairs found in systems where a first-order resonant pair has already been detected. Out of 187 possible same-system second pairs accompanying a first-order resonant pair, 61 were also near a first-order resonance. Therefore, the conditional probability of detecting a second first-order resonant pair after detecting the first one is 61/187 or 32.6$\pm3.4$\%. This is twice as high as the raw incidence of first-order resonances, 15.6$\pm1.1\%$.

To further test this result, we carried out a bootstrap experiment. We generated 100 samples of synthetic multi-planet ($\geq 3$-planet) systems for which
there was, by design, no correlation in the occurrence of resonances within the same system. 
We kept the total number of systems (262) and the planet multiplicity in each system the same as in the observed sample. However, orbital period ratios were randomly drawn from the global period ratio distribution. We computed the same set of resonant statistics for these synthetic samples as we did for the observed sample. The results are shown on the right panel of Fig. \ref{fig:covariance}. The simulated systems have significantly fewer cases where two first-order resonant pairs appeared in the same system (11, down from 61). We performed a contingency table test as implemented in {\tt scipy.chi2\_contingency}. For systems with an existing first-order resonant pair, it is indeed much more likely to find another resonant pair with a p-value of order 10$^{-7}$. For systems with an existing second-order resonant pair, it is also more likely to see another resonant pair, although the statistical significance is weaker (p-value = 0.048). This is perhaps because second-order resonance is intrinsically rare: there is a much smaller sample with which we can perform this test (second row of Fig
. \ref{fig:covariance}). In other words, resonant pairs tend to congregate within systems. See also relevant results reported by \citet{Jiang2020}.

\begin{figure}
\center
\includegraphics[width = 1.\columnwidth]{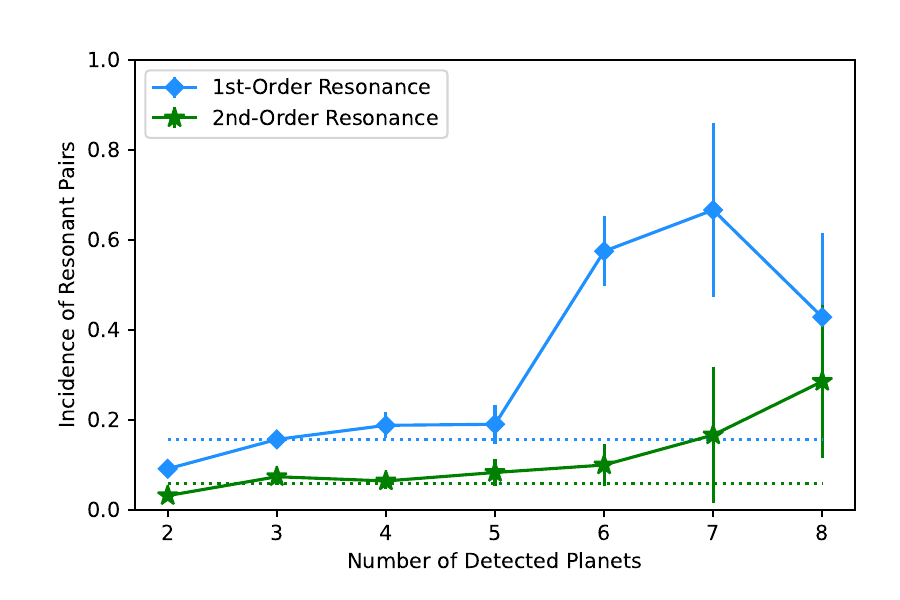}
\caption{The incidence of 1st and 2nd-order resonant pairs increases with the observed planet multiplicity. For the 1st-order resonances, the incidence within systems of two detected planets is 9.1$\pm1.3$\% (43/470). It steadily increases to 15.7$\pm2.0$\% (53/338) for 3-planet systems, 18.8$\pm2.9$\% (35/186) for 4-planet systems,  19.0$\pm4.3$\% (16/84) for 5 planet systems, and 57.5$\pm7.8$\% (23/40) for 6-planet systems. We caution there is only one 7-planet system \citep[TRAPPIST-1,][]{Gillon,Luger2017} and one 8-planet system \citep[Kepler-90,][]{Shallue} in our sample. Both systems evince an abundance of resonances. The horizontal dotted lines are the incidence of resonant pairs across all multi-planet systems irrespective of age. }
\label{fig:n_p_resonance}
\end{figure}

What kind of planetary systems are more likely to host resonant pairs? We found that the incidence of resonant pairs steadily increases with the number of detected planets in each system (Fig. \ref{fig:n_p_resonance}). This trend holds even after accounting for the fact that systems with higher planet multiplicity have more planet pairs. For first-order resonances, the incidence within systems of two detected planets is 9.1$\pm1.3$\% (43/470). It steadily increases to 15.7$\pm2.0$\% (53/338) for 3-planet systems, 18.8$\pm2.9$\% (35/186) for 4-planet systems,  19.0$\pm4.3$\% (16/84) for 5 planet systems, and 57.5$\pm7.8$\% (23/40) for 6-planet systems. The incidence of second-order resonances follows a similar trend (Fig. \ref{fig:n_p_resonance}). The incidence increases from 15/470 (3.2$\pm0.8$\%) for 2-planet systems to 4/40 (10$\pm5$\%) for 6-planet systems.


The number of detected transiting planets in a planetary system bears on the mutual inclinations in that system \citep{Fabrycky2014}. If the mutual inclinations are large, a smaller fraction of the true number of planets will be detected by the transit method because their orbital planes are less likely to simultaneously intersect the line of sight with the star \citep{Zhu2018,He2020, Millholland2021}. Resonant systems formed by Type I migration are expected to have small mutual inclinations as a consequence of inclination damping in the disk \citep{Izidoro}. Indeed, TRAPPIST-1 is an example of a high-multiplicity, low-inclination resonant chain system \citep{Gillon,Luger2017}. \citet{Agol2021} found the seven TRAPPIST-1 planets to have mutual inclinations of order 0.04$^\circ$. In contrast, the {\it Kepler} planets in general have mutual inclinations of order 1-2$^\circ$ and are predominantly non-resonant \citep{Fabrycky2014}. Our results suggest that planets in high-planet-multiplicity, low-mutual-inclination systems are also more likely to be near-resonant. Perhaps the dynamical evolution in some planetary systems is especially quiescent, preserving both the resonances and the low mutual inclinations \citep[see also simulations by][]{Dawson2016d}. In systems that today are not near resonance, the same mechanism responsible for breaking the early resonances, such as a violent dynamical instability, may have also generated a larger range of mutual inclinations \citep{Izidoro,Esteves2020}.

One requirement for successful capture into resonance is adiabaticity: disk-driven migration must occur over a timescale longer than the resonant libration period.  Our finding that the majority of young, close-in planets have near-resonant orbits (Section \ref{sec:young_resonance}) may favor slower disk migration and lower disk turbulence. Stochastic forces due to disk turbulence can dislodge planets from resonance \citep[e.g.][]{Rein2009,Batygin_turbulence,Wu_Chen_Lin2024}. The rate of disk migration depends on several factors such as disk surface density, radial profile of the disk, viscosity among those. \citet {Lee2016} suggested the characteristically low gas-mass fractions of sub-Neptunes, and the need to shed most of the disk gas to allow solid proto-cores to dynamically excite each other and merge into super-Earth cores, also point to 
gas-poor (but not gas-empty) disks as the birth environments for {\it Kepler} planets. Disk migration is naturally slow during this last phase of gas disk dispersal.

\subsection{Resonances are Common among Young Planets}\label{sec:young_resonance}

Are resonances more common among younger planetary systems? Among the 7 multi-planet systems younger than 100 Myrs, 6/7 (86$\pm13$\%) show at least one near-resonant (first or second-order) planet pair. The only exception is TOI-942, and even in that case, there is a hint of a possible resonance. The period ratio of TOI-942 b and c is very close to $1.5^2$. If an undetected planet were placed at the geometric center, it would be in a 3:2 resonance with both known planets, and the average $\Delta$ could be as low as 0.021. TOI-942 has only been observed in {\it TESS} Sector 5, and there are too few transits for a TTV analysis. We discuss some caveats due to selection bias in Section \ref{sec:caveats}.

Moving onto the pairwise statistics among young $<100$-Myr planets, we find that 7/10 (70$\pm15$\%) adjacent pairs are close to first-order resonances. This is significantly higher than the 14.7$\pm1.8$\% observed in mature systems ($>$1-Gyr-old). The 3:2  is the most common resonance seen in young planetary systems, accounting for 6/10 of the detected resonances. As noted at the beginning of this section, the 3:2 seems to be the most prevalent across all ages. Detailed statistics are provided in Table~\ref{tab:mmr}. 

 The incidence of resonance among adolescent systems (0.1 to 1-Gyr-old) lies between those of the young and mature populations: 6/16 
 (38$\pm$12\%) of adolescent systems contain at least one resonant pair, compared to 6/7 (86$\pm$13\%) for the young bin, and 58/255 (22.7$\pm$2.6\%) for the mature bin.  
 This trend is echoed when counting pairs: 7/29 ($24.1\pm7.9$\%) of the neighboring adolescent pairs are close to first-order resonances, lower than the 7/10 (70$\pm15$\%) observed in young pairs, and higher than the 59/402 (14.7$\pm1.8$\%) observed in mature pairs.
 
\begin{figure}
\center 
\includegraphics[width = 1.\columnwidth]{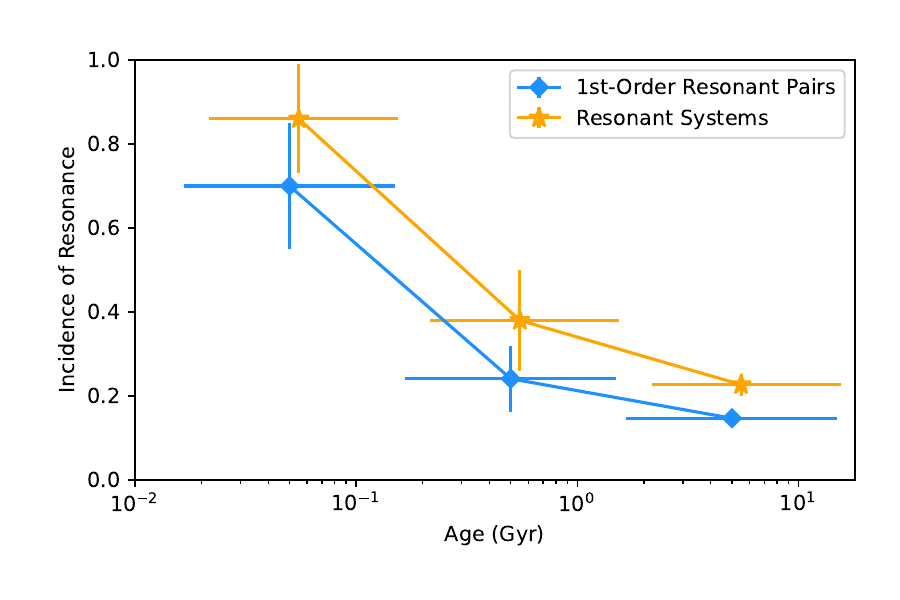}
\includegraphics[width = 1.\columnwidth]{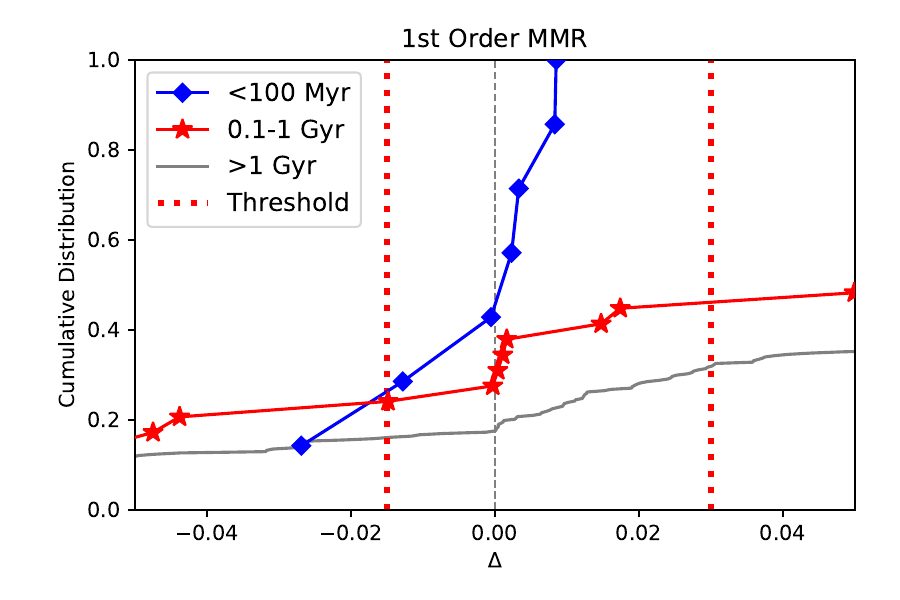}
\caption{{\bf Top:} The incidence of 1st-order resonant pairs (blue curve) and the incidence of planetary systems with at least one resonant pair (1st or 2nd-order; orange curve) across three age bins: (1) `young' ($<$ 100 Myr), (2) `adolescent' (100-1000 Myr), and  (3) `mature'
 ($>$ 1 Gyr). The young planetary systems abound in resonances, which appear to disintegrate over a timescale of order 100 Myr. The age bins have been offset for clarity. 
 {\bf Bottom:} Cumulative distributions of $\Delta$ near 1st-order commensurability across different age groups. Data are plotted for all pairs resonant and non-resonant; the red and black curves saturate  at large $\Delta$ values not plotted. We see evidence of the $\Delta$ distribution broadening with time, particularly within the first 1 Gyr.}
\label{fig:age_resonance}
\end{figure}

 Fig.~\ref{fig:age_resonance} summarizes the incidence of resonant configurations as a function of age as well as a cumulative distribution of $\Delta$ near first-order resonance. Our results indicate that resonant configurations are practically universal among young $< 100$-Myr planetary systems, and that the resonances dissolve over timescales on the order of 100 Myr.   An Anderson-Darling test using {\tt anderson\_ksamp} in {\tt scipy} suggests that the $\Delta$ distribution of young $< 100$-Myr planetary systems is markedly distinct from the mature sample ($>1$-Gyr) with a p value of 0.0025.  These findings directly support the `breaking-the-chains' model \citep{Izidoro,Izidoro2021,Goldberg2022}, in which close-in planetary systems are initially assembled into resonant configurations that are gradually disrupted by dynamical instabilities.

Within the Solar System, there is evidence for dynamical instabilities and rearrangement within the first 100 Myr. For example, the giant impacts phase of terrestrial planet formation, including the Moon-forming impact, is suspected to have occurred within the first tens of Myr after gas disk dissipation \citep[e.g.][]{Chambers2013,Borg}. 
In the Nice model, an initially resonant configuration of the Solar System's giant planets destabilized within 100 Myr after their formation \citep{Tsiganis,Nesvorny2018}. Similar dynamical upheavals on similar timescales in exoplanet systems can explain the decline of resonant configurations that we observe. 




\begin{deluxetable*}{|l|l|l|l|l|l|}
\tablecaption{Prevalence of Resonant Configuration Across Ages \label{tab:mmr}} 
\tablehead{
\colhead{} & \colhead{$<100$Myr} & \colhead{100Myr-1Gyr} & \colhead{$>1$Gyr}& \colhead{All Ages $^4$} }
\startdata
\hline
First-order Resonant Pairs &7/10 (70$\pm15$\%)&7/29 (24.1$\pm7.9
$\%)&59/402 ($ 14.7\pm 1.8 \%$) & 177/1131 (15.6$\pm1.1\%$)\\
\hline
2:1 ($-0.015\leq \Delta \leq 0.03$)& 1/10 (10$\pm10$\%) &  2/29 (6.9$\pm4.7$\%)& 15/402 (3.7$\pm0.9$\%) & 57/1131 (5.0$\pm0.7\%$)\\
\hline
3:2 ($-0.015 \leq \Delta \leq 0.03$)& 6/10 (60$\pm16$\%) &  5/29 (17.2$\pm7.0$\%)& 30/402 (7.5$\pm$1.3\%) & 83/1131  (7.3$\pm0.8\%$)\\
\hline
4:3 ($-0.015 \leq \Delta \leq 0.03$) & 0/10 &  0/29 & 8/402 (2.0$\pm0.7$\%)  & 23/1131  (2.0$\pm0.4\%$)\\
\hline
5:4 ($-0.015 \leq \Delta \leq 0.03$) & 0/10  &  0/29& 5/402 (1.2$\pm0.6$\%)  & 12/1131  (1.1$\pm0.3\%$)\\
\hline
6:5 ($-0.015 \leq \Delta \leq 0.03$)& 0/10 &  0/29& 1/402 (0.2$\pm0.2\%$)  & 2/1131  (0.2$\pm0.1\%$)\\
\hline
&&&\\
\hline 
Second-order Resonant Pairs & 1/10 (10$\pm10$\%)& 4/29 (13.8$\pm6.4\%$)& 23/402 (5.7$\pm1.2$\%) &66/1131  (5.8$\pm0.7\%$)\\
\hline
3:1 ($-0.015 \leq \Delta \leq 0.015$)& 0/10 &  0/29& 2/402 (0.5$\pm0.4$\%)  & 9/1131  (0.8$\pm0.3\%$)\\
\hline 
5:3 ($-0.015 \leq \Delta \leq 0.015$)& 1/10  (10$\pm10$\%)&  3/29 (10.3$\pm5.7$\%)& 11/402 (2.7$\pm0.8$\%) & 33/1131  (2.9$\pm0.5\%$)\\
\hline
7:5 ($-0.015\leq \Delta \leq 0.015$)& 0/10 &  1/29 (3.4$\pm3.4$\%)& 6/402 (1.5$\pm0.4$\%) & 17/1131  (1.5$\pm0.6\%$)\\
\hline
9:7 ($-0.015 \leq \Delta \leq 0.015$)& 0/10& 0/29& 4/402 (1.0$\pm0.5$\%) & 7/1131  (0.6$\pm0.2\%$)\\
\hline 
Incidence of Resonant Pairs& 8/10 (80$\pm13$\%) &  11/29 (37.9$\pm9.0$\%)& 82/402 (20.4$\pm2.0$\%) & 243/1131 (21.5$\pm1.2\%$)\\
\hline 
Incidence of Resonant Systems $^1$& 6/7 (86$\pm13$\%) &  6/16 (38$\pm12$\%)& 58/255 (22.7$\pm2.6$\%) & 182/734 (24.8$\pm1.6\%$)\\
\hline
Multi-Resonant Systems $^2$ & 2/2 (100$\pm70$\%) &  2/9 (22$\pm14$\%)& 12/89 (13.5$\pm$ 3.6\%) & 37/262 (14.1$\pm2.2\%$)\\
\hline
Resonant Chains $^3$ & 2/2 (100$\pm70$\%) &  2/9 (22$\pm14$\%)& 10/89 (11.2$\pm$ 3.3\%) & 30/262 (11.5$\pm2.0\%$)\\
\hline
\enddata
\tablecomments{These are raw fractions uncorrected for detection biases. 1. Resonant systems are planetary systems containing at least one resonant pair (1st or 2nd-order). 2. Multi-resonant systems are systems with two or more resonant pairs (can be non-contiguous). 3. Resonant chains are systems that contain two or more contiguous resonant pairs (e.g. a 3-planet system with period ratios 2:3:5). 4. Most of the `All Ages' sample do not have reported ages.}
\end{deluxetable*}

\subsection{Planets in the Radius Gap are Least Likely to be Resonant}

\begin{figure*}
\center
\includegraphics[width = 1.5\columnwidth]{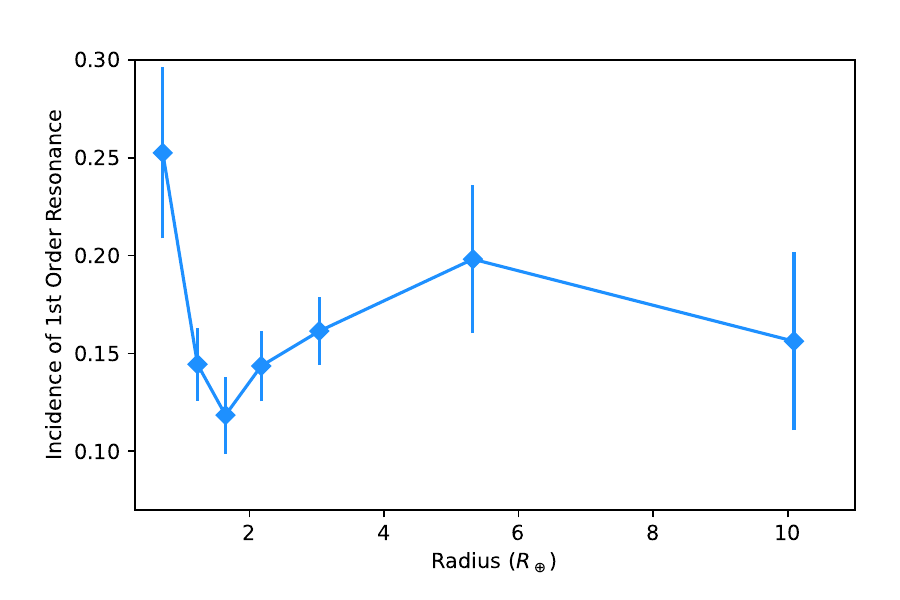}
\caption{The incidence of adjacent pairs near first-order resonances ($-0.015\leq \Delta \leq 0.03$) as a function of planet radii. The boundaries of the radius bins are strategically placed at 1) Earths and sub-Earths ($R_p<1R_\oplus$; 25/99; 25.3$\pm4.4\%$); 2) super-Earths ($1R_\oplus \leq R_p<1.5R_\oplus$; 51/353; 14.4$\pm1.9\%$); 3) planets in the radius gap identified by \citet{Fulton} ($1.5R_\oplus\leq R_p<1.9R_\oplus$; 32/270; 11.9$\pm2.0\%$); 4) sub-Neptunes ($1.9R_\oplus \leq R_p<2.5R_\oplus$; 54/376; 14.4$\pm1.8\%$); 5) Neptunes ($2.5R_\oplus \leq R_p<4R_\oplus$; 72/446; 16.1$\pm1.7\%$); 6) sub-Saturns ($4R_\oplus \leq R_p<8R_\oplus$; 22/111; 19.8$\pm3.8\%$); 7) Jupiters ($R_p \geq 8R_\oplus$; 10/64; 15.6$\pm4.5\%$).  Planets within the radius gap have the lowest chance of being near first-order resonances.}
\label{fig:rp_resonance}
\end{figure*}

Does the likelihood of resonance depend on the size of the planets? 
Fig.~\ref{fig:rp_resonance} shows the incidence of 1st-order resonance as a function of planet radius, 
across all ages (known and unknown). Incidence is calculated on a per-planet basis. A planet is considered near-resonant if its interior or exterior pair is near-resonant.  We did not adjust our standard bounds on $\Delta$ for 1st-order resonance ($-0.015\leq \Delta \leq 0.03$) to account for how libration widths change with planet-to-star mass ratio; planet masses are largely unmeasured, as are the eccentricities \citep[e.g.][]{Xie,vaneylen}, which also affect libration widths. Our bounds on $\Delta$ are likely wide enough to include resonant planets of varying radii. For example,  \citet{Millholland876} reported $\Delta\approx1.5\%$ for the giant planets of the GJ-876 system \citep{Rivera2005}.


The radius bins in Fig.~\ref{fig:rp_resonance}, and their corresponding 1st-order resonance fractions, are as follows: 1) Earths and sub-Earths ($R_p<1R_\oplus$; 25/99; 25.3$\pm4.4\%$); 2) super-Earths ($1R_\oplus \leq R_p<1.5R_\oplus$; 51/353; 14.4$\pm1.9\%$); 3) planets in the radius gap identified by \citet{Fulton} ($1.5R_\oplus\leq R_p<1.9R_\oplus$; 32/270; 11.9$\pm2.0\%$); 4) mini-Neptunes ($1.9R_\oplus \leq R_p<2.5R_\oplus$; 54/376; 14.4$\pm1.8\%$); 5) Neptunes ($2.5R_\oplus \leq R_p<4R_\oplus$; 72/446; 16.1$\pm1.7\%$); 6) sub-Saturns ($4R_\oplus \leq R_p<8R_\oplus$; 22/111; 19.8$\pm3.8\%$); 7) Jupiters ($R_p \geq 8R_\oplus$; 10/64; 15.6$\pm4.5\%$). We capped the `radius-gap' planets at 1.9 $R_\oplus$ because recent works \citep[e.g.][]{Rubenzahl1347} have shown that 1.9 $R_\oplus$ or 10 $M_\oplus$ represents the maximum size of rocky planets with minimal H/He envelopes \citep[see also][]{Hu2024}.

Super-Earths exhibit the lowest
incidence of first-order resonances, especially those within the radius gap ($1.5R_\oplus\leq R_p<1.9R_\oplus$; 32/270; 11.9$\pm2.0\%$). Gilbert and Petigura (in review) have reported that planets within the radius gap tend to have larger orbital eccentricities than planets with either smaller or larger radii. These findings together suggest that the radius gap may be populated by planets that have undergone dynamical instabilities such as giant impacts. These instabilities could have excited orbital eccentricites, disrupted resonances, and increased planet masses via collisions.

 Progressively higher incidences of resonances are seen for mini-Neptunes ($1.9R_\oplus \leq R_p<2.5R_\oplus$; 54/376; 14.4$\pm1.8\%$), Neptunes ($2.5R_\oplus \leq R_p<4R_\oplus$; 72/446; 16.1$\pm1.7\%$), and sub-Saturns ($4R_\oplus \leq R_p<8R_\oplus$; 22/111; 19.8$\pm3.8\%$). \citet{Millholland2019} reported a related trend, noting that near-resonant planets tend to have larger radii than non-resonant planets, and proposed that obliquity tidal heating of near-resonant planets might be responsible for inflating their radii. Another possibility is that larger planets formed farther out in the parent disk, where cooler temperatures and lower opacities facilitated the accretion of voluminous gas envelopes; puffier planets would thus have migrated larger distances to arrive at their current locations, and had more opportunity to capture into resonance   \citep{Lee2016,Lee_accretion}. Larger planets may also be more massive, and more massive planets capture more easily into resonance \citep[e.g.][]{Batygin_Petit}.
 


Fig.~\ref{fig:rp_resonance} shows that the observed incidence of resonance for Jupiter-sized planets is lower than that of sub-Neptunes or sub-Saturns ($R_p \geq 8R_\oplus$; 10/64; 15.6$\pm4.5\%$). 
Our low resonance fraction for Jupiters is in tension with \citet{Wright2011} who found that $\sim$1/3 of giant planets discovered by the radial-velocity (RV) method could be near-resonant, based on their observed periods. The discrepancy in occurrence rate may largely be due to how our respective samples differ: we select for transiting planets and exclude RV planets. 
Our transiting giant planet sample contains many hot Jupiters whose formation/migration pathway likely involves the disruption of resonances with other planetary companions (\citealt{Knutson2014}; \citealt{Dawson}).


\subsection{Are Resonant Chains Parked at Disk Inner Edges?}\label{sec:disk_edge}

 It has been suggested that disk inner edges play a pivotal role in forming resonant chains \citep[e.g.][]{Masset_2006,Ormel2017,Izidoro,Ogihara2018,Dai1136,Wong2024}. In general, capture into resonance requires convergent migration, i.e., period ratios have to decrease \citep[e.g.][]{Batygin2015_capture}. To ensure convergent encounters when all planets are migrating inward, exterior planets must migrate faster than interior planets, a condition not necessarily met in protoplanetary disks \citep[e.g.][]{Tamayo2017,Wong2024}. 

A planet can stop migrating inwards when it reaches the disk's inner edge  \citep{Masset_2006}. There it can wait as inwardly migrating exterior planets catch up. This scenario ensures convergent encounters, and allows for the construction of a resonant chain as successive convergent encounters lead to resonance captures---presuming the inner planet can act as an anchor, and is not torqued further inward by the chain and the surrounding disk (\citealt{AtaieeKley21}).

The inner edges of protoplanetary disks are commonly thought to be set by magnetospheric truncation---where the magnetic pressure exerted by the host pre-main-sequence star disrupts the disk accretion flow (e.g.~\citealt{ShapiroTeukolsky1986}). The theory of ``disk locking'' further posits that material at this truncation radius co-rotates with the star (e.g.~\citealt{GhoshLamb1979}; \citealt{Konigl1991}; \citealt{OstrikerShu1995}; \citealt{Long, Romanova}). The light curves of young ``dipper stars'' appear to trace disk material rotating with the same periods as their stars (e.g.~\citealt{Stauffer2015}). \citet{Batygin_innerdisk} further suggested that the inner disk edge has a similar Keplerian period of about 2-12 days across a wide range of host star masses. 



\begin{figure*}
\center
\includegraphics[width = 1.4\columnwidth]{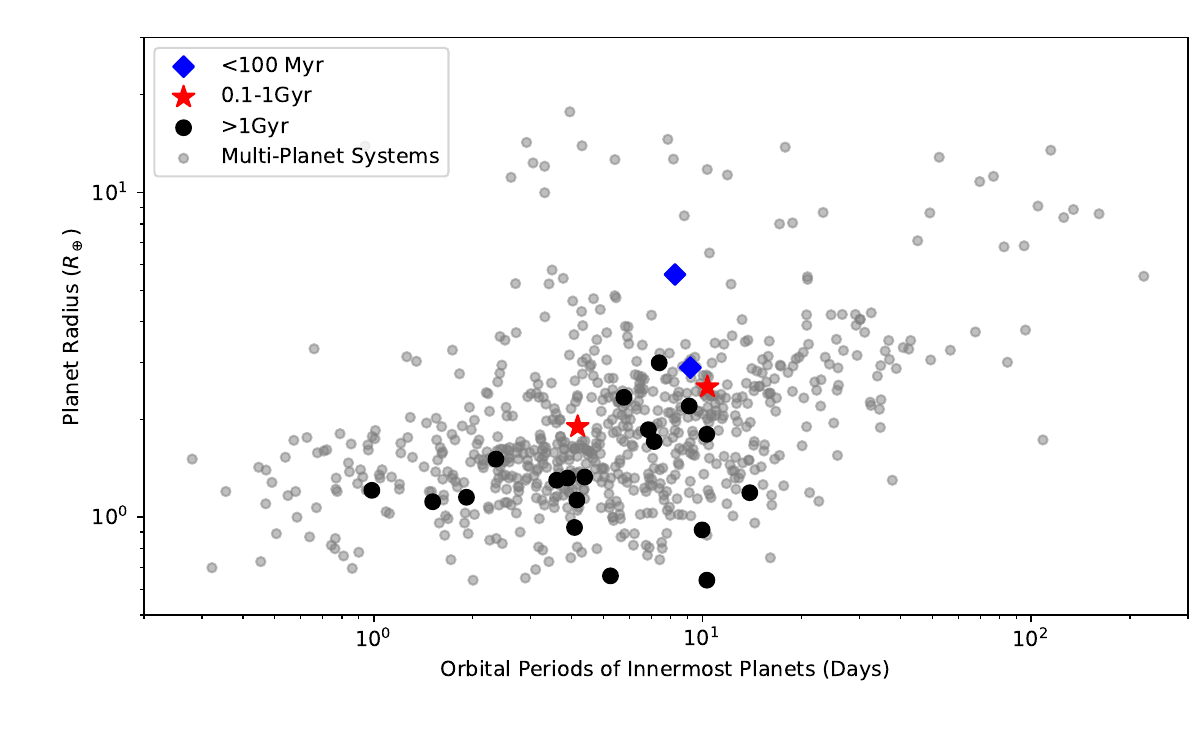}
\includegraphics[width = 1.4\columnwidth]{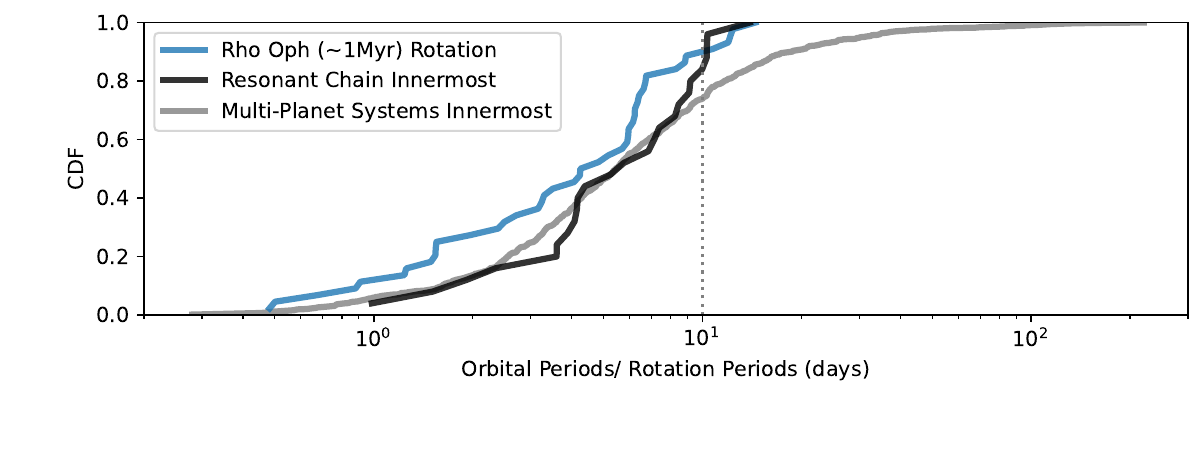}
\caption{The orbital periods and radii of the innermost planets in multi-planet systems. The resonant chain systems (solid symbols, systems with unbroken multiple resonant pairs) and generic multi-planet systems (gray circles) are shown separately. The bottom panel is the rotation period distribution of young stars in rho Ophiuchus \citep[$\sim$1-Myr-old][]{Rebull2018} which is still in the disk-hosting stage. If resonant chains are parked near the disk inner edges, the orbital period distribution of the innermost planets should trace the rotation period distribution of young stars. Indeed, almost all resonant chain systems have innermost planets $<$10 days, the same applies to the rotation periods of young stars. On the other hand, 25$\pm2\%$ of generic multi-planet systems have their innermost planets between 10-100 days.}
\label{fig:innermost}
\end{figure*}

\citet{Lee_usp} explained how the occurrence rate profiles of {\it Kepler} planets (as measured by, e.g., \citealt{Fressin,Dressing_mdwarf,CKS4}) can be understood in terms of the inner disk edge: exterior to the disk edge, in the disk proper, planets are abundant, whereas interior to the disk edge, in the magnetospheric cavity, planets are rare. Under the assumption of disk locking, \citet{Lee_usp} utilized the observed rotation periods of pre-main-sequence stars in stellar clusters to establish that  orbital periods at the inner disk edge range from $\sim$0.3--20 days.  Here we follow their reasoning and compare, in Figure~\ref{fig:innermost}, the stellar rotation periods in the $\sim$1-Myr-old cluster rho Ophiuchus \citep{Rebull2018}, to the orbital periods of the innermost planets in resonant chains. There is generally good overlap. Within rho Oph, $\sim$94\% of stars rotate faster than 10 days, and the longest rotation period is $\sim$14 days. Among the innermost planets of resonant chains, $\sim$90\% have orbital periods shorter than 10 days, and the longest period is that of Kepler-127 at 14.4 days. We contrast these distributions with the period distribution of the innermost planets of generic multi-planet systems (gray points in Fig.~\ref{fig:innermost}). The latter exhibit a broader range of periods: e.g., 14.5\% lie outside 15 days, with some exceeding 100 days. There are 30 resonant-chain systems known; if their innermost planets were distributed like those in generic multi-planet systems, we would expect, in a sample size of 30, an average of 4 planets to have orbital periods $> 15$ days. Yet none are observed.




While we have shown that the observed orbital periods of resonant chain systems are consistent with being parked at inner disk edges (e.g.~\citealt{TerquemPapaloizou2007}; 
\citealt{Wong2024}), there remain unresolved issues. Observationally, transit surveys are biased toward finding shorter-period planets \citep{Winn2010}, and we may be missing resonant chains parked farther out. \citet{Miranda} suggested that planets may halt their migration farther out from the inner disk radius, by a factor of 3--5, if the Type-I migration torque is calculated including wave reflections at the inner edge. At these close-in distances, the disk may be magnetically turbulent (e.g. \citealt{DeschTurner2015,Jankovic2021,Jankovic2022}), and stochastic forcing may prevent planets from parking \citep{Wu_Chen_Lin2024}. Even if the innermost planet parks initially, continued angular momentum transfers up and down the  chain, and with the disk, may push the inner planet further in, depending on the details of disk transport (\citealt{AtaieeKley21}). Staging the creation of chains at the tail end of the disk's life, when gas surface densities are low (e.g.~\citealt{Lee2016}), may alleviate these concerns.

\subsection{Resonant Repulsion}
\label{sec:repulsion}

`Resonant repulsion' refers to the process where the deviation from perfect period commensurability $\Delta$ near first-order resonance increases over time due to some dissipative process \citep{Papaloizou2010,Lithwick_repulsion,Batygin_repulsion,Delisle2014}. Eccentricity tides raised on the planet by the star provide one source of dissipation. 
Under the assumption that tidal dissipation in the planet drives resonant repulsion, and that a pair starts at exact period commensurability (i.e. $\Delta=0$ at $t=0$),
\citet{Lithwick_repulsion} derived that $\Delta$ evolves as
\begin{equation}\label{eqn:repulsion}
\begin{aligned}
& \  \Delta(t) \approx 0.006 \left( \frac{Q_{\rm p}^\prime}{100} \right)^{-1/3} \left( \frac{M_{\rm p}}{10M_\oplus} \right)^{1/3} \left( \frac{R_{\rm p}}{2R_\oplus} \right)^{5/3}\\
 &\quad \times \left( \frac{M_\star}{M_\odot} \right)^{-8/3} \left( \frac{P_{\rm p}}{5 \,{\rm days}} \right)^{-13/9} \left( \frac{t}{5 \,{\rm Gyr}} \right)^{1/3}\\
 &\quad \times (2\beta+2\beta^2)^{1/3}
\end{aligned}
\end{equation}
where $Q_{\rm p}^\prime$ is the reduced tidal quality factor of the inner planet; $M_p$, $R_p$, and $P_{\rm orb}$ are the mass, radius, and orbital period of the inner planet;  $M_\star$ is the mass of the host star; and $\beta \equiv M_{\rm out}\sqrt{a_{\rm out}}/(M_{\rm p}\sqrt{a_{\rm p}})$ contains the mass and semi-major axis ratio of the outer and inner members of the pair. We note $\Delta=0$ at $t=0$ is a simplification. The initial $\Delta$ can have a range of non-zero values when the resonant chain is first established, when migration and eccentricity damping are driven by the parent disk (see, e.g., \citealt{TerquemPapaloizou2019,Choksi2020,Dai1136}). 

\begin{figure*}
\center
\includegraphics[width = 2.\columnwidth]{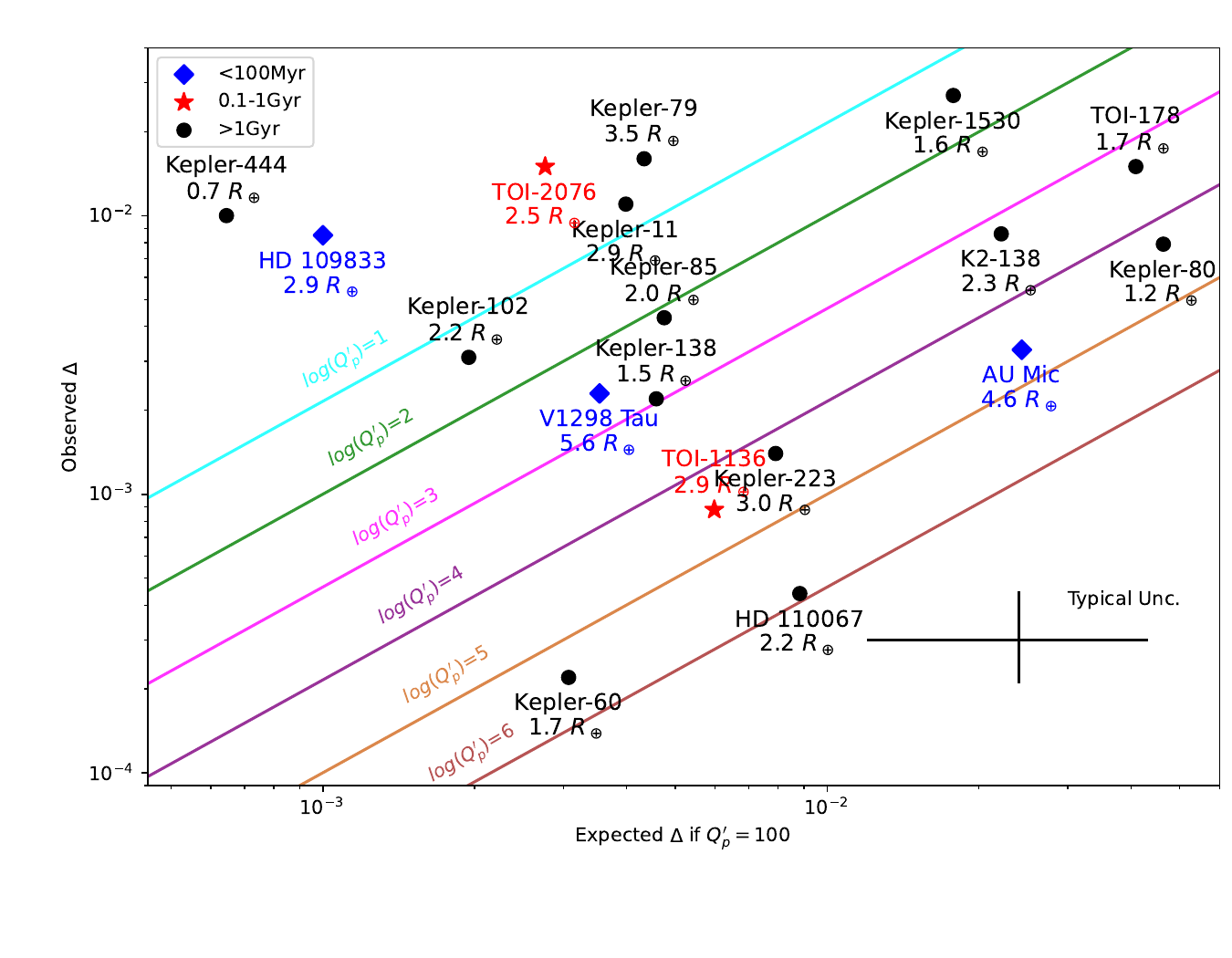}
\caption{The observed deviation $\Delta$ for the resonant-chain systems with well-measured ages vs. the cumulative eccentricity damping, i.e.~all terms in Eqn.~\ref{eqn:repulsion} except the reduced tidal quality factor. The solid lines shows the theoretically predicted $\Delta$ at various reduced tidal quality factors. We have labelled each resonant-chain system with the radius of the planet that is expected to experience the strongest tides.  TOI-1136 \citep[700-Myr-old;][]{Dai1136} and AU Mic \citep[20-Myr-old;][]{Plavchan} are still deep in resonances ($\Delta\sim0.1\%$), and the inferred reduced tidal quality factor is roughly $10^4-10^5$, which is consistent with Uranus \citep[e.g.][]{Gomes2024}. However,
HD 109833 \citep[10 to 16-Myr-old;][]{Wood}, TOI-2076 \citep[200-Myr-old;]{Hedges} are substantially far away from resonance ($\Delta\sim1\%$) that the required reduced tidal quality factor $Q_p^\prime<10$ seems unphysical. Other processes such as disk e-damping \citep{Choksi2020}, disk turbulence \citep{Goldberg2023}, obliquity tides \citep{Millholland_obliquity}, planetesimal scattering \citep{Chatterjee_2015} should be included to explain the observed $\Delta$ distribution.}
\label{fig:repulsion}
\end{figure*}

Using our resonant-chain systems with well-measured ages together with eqn.~\ref{eqn:repulsion}, one can constrain the reduced tidal quality factor $Q'_{\rm p}$ of planets. 
We computed the expected resonant repulsion for each planet in the resonant-chain systems. We used the median value for planet mass from the literature, if available. If mass was not available, we estimated the mass using the reported radius and the mass-radius relationship from \citet{Chen_Kipping}. We selected the planets that experience the strongest dissipation in each system. This is usually the innermost planet unless an outer planet has a much larger radius. Previous works have shown that tidal dissipation on one planet in a resonant chain can affect other planets in the chain \citep[e.g.][]{Papaloizou2018,Brasser2021,Dai1136}.  Fig. \ref{fig:repulsion} shows the observed $\Delta$ vs. the expected $\Delta$ as computed by eqn. \ref{eqn:repulsion} with different $Q'_{\rm p}$. 

Among the young systems, TOI-1136 \citep[700-Myr-old;][]{Dai1136} and AU Mic \citep[20-Myr-old;][]{Plavchan} are relatively deep in resonance ($\Delta\lesssim0.1\%$). The inferred reduced tidal quality factors are roughly $10^3-10^4$. Given their Neptune-like and sub-Neptune radii, such a $Q_p^\prime$ is not surprising as it is close to that of Uranus \citep[see recent result by][]{Gomes2024}. However,
HD 109833 \citep[10 to 16-Myr-old;][]{Wood} and TOI-2076 \citep[200-Myr-old;][]{Hedges} are significantly farther from commensurability ($\Delta\sim1\%$), such that the required $Q'_{\rm p}$ is smaller than 1. This demands an implausibly fast dissipation rate, especially considering the planets are sub-Neptune rather than terrestrial in size. Therefore, resonant repulsion from equilibrium eccentricity tides cannot be the full story here.

A number of previous works have noted the same issue: the reduced tidal quality factor required to explain the observed $\Delta$ distribution of Kepler planets is too small, even for systems assumed to be of solar age \citep[e.g.,][]{LeeMH,Silburt}. As a possible solution, \citet{Millholland_obliquity} invoked obliquity tides, where near-resonant planets are also locked in a high-obliquity Cassini state. With obliquity tides, near-resonant planets experience significantly enhanced tidal dissipation. However, this mechanism does not seem compatible with the non-zero free eccentricities observed in many near-resonant systems, which should have been damped out by tides, whether obliquity or eccentricity in origin \citep{Choksi2023,Goldberg2023}.

Near-resonant planets must have attained substantial deviations from commensurability $\Delta$ either before gas disk dispersal or shortly after. For instance, eccentricity damping due to disk-planet interaction can initially capture planets wide of resonance \citep{TerquemPapaloizou2019, Choksi2020, Dai1136}. The magnitude of $\Delta \sim +1\%$ can be explained if the ratio of the migration timescale to eccentricity damping timescale, which is set by the local disk structure, is $\sim$$10^4$. Stochastic forcing generated by disk turbulence \citep{Rein2012,Goldberg2023,Wu_Chen_Lin2024} can also produce a wider distribution of $\Delta$, though possibly too wide: the trough at $\Delta < 0$  may be over-filled, contrary to observation (see Section \ref{sec:ident} and Figure \ref{fig:delta}). Alternatively, one-sided torques during expansion of the inner magnetospheric cavity of the disk might pull planets out of resonance shortly before the disk dispersal \citep{Liu2017,Hansen2024}. Finally,  gravitational scattering by planetesimals and planet embryos have been argued to reproduce the observed distributions of $\Delta$ and free eccentricity \citep{Chatterjee_2015,Raymond2021,Wu2024}. More detailed investigations are required to determine whether these mechanisms can indeed operate on the correct timescale, and in the specific young planetary systems that have been discovered so far. 







\section{Caveats and Future Work}
\label{sec:caveats}
We point out a few shortcomings of our work and avenues for future studies:

\begin{itemize}
    \item {We have only examined 2-body resonances, but 3-body resonances can exist and have been observed \citep[e.g., Kepler-221,][]{Fabrycky2014,MacDonald2016,Goldberg2021}. Assuming there was an unbroken 3-body resonance, \citet{Luger2017} were able to discover the seventh planet in TRAPPIST-1 in the low signal-to-noise ratio K2 light curve. See also HD 110067 \citep{Luque}.}

    \item{Our analysis is mostly based on the observed period ratios. We are unable to distinguish between resonant (librating) from near-resonant (circulating) solutions. A proper TTV or photodynamical modeling \citep[e.g.][]{Carter,Mills2016} is required to elucidate the dynamics of young multi-planet systems. Such an analysis is generally not feasible on {\it TESS}-discovered planets due to the short observational baseline. However, demanding long-term orbital stability can sometimes help pin down the orbital architecture of a multi-planet system \citep{Fabrycky8799,Wang_8799,Tamayo2017,Lammers}.}

\item{We have only computed the raw incidence of resonances (fraction of planet pairs near a period commensurability over the total number of pairs), and have not corrected for detection biases. Missing planets could alter the observed period ratio distribution. To uncover the complete inventory of planets in young planetary systems, both transit and radial velocity follow-up observations are necessary. However, radial velocity follow-up is particularly challenging for young planetary systems due to elevated stellar activity \citep[see e.g.][]{Blunt}.}

      \item {It may be the case that previous works only reported young planetary systems that are near-resonant because they could be confirmed by the detection of TTVs. This bias could cause us to overestimate the prevalence of resonances among young systems. However, this seems unlikely because most of the reported young systems were discovered after just one or two sectors of {\it TESS} observations (28-days in duration) or a single K2 campaign (80-days). TTVs are difficult to detect on such short baselines. Most of the time, TTVs were only detected after intensive ground-based follow-up observation. For example, the AU Mic discovery paper did not mention TTVs at all; TTVs were reported in follow-up papers \citep{Plavchan,Wittrock}. In fact, the presence of TTVs actually lowers the detectability of transiting planets in a traditional box-least-square search \citep{Kovac2002}. This effect disfavors the detection of young resonant systems.}

      \item{Our analysis has focused on MMRs which are perhaps the most delicate feature of planetary orbital architecture. MMR may be easily disrupted by dynamical evolution \citep{Izidoro,Goldberg_stability}. We encourage future work to investigate planetary multiplicity, mutual inclination, and orbital eccentricities as a function of age to provide a more comprehensive understanding of the dynamical evolution of planetary systems.}

      \item{Galactic evolution sets the backdrop of planet formation particularly in terms of overall metallicity, star formation rate, kinematics due to mergers \citep{Matteucci,Bland-Hawthorn}. However, galactic evolution unfolds on many-Gyr timescales. Our young ($<100$-Myr-old) and adolescent ($0.1-1$-Gyr-old) are essentially indistinguishable on a such a timescale. Future works on an `old' ($>10$-Gyr-old) sample of planetary systems should be mindful of the influence of galactic evolution on planet formation.}

      \item{Although we identified trends in resonance fraction with age (Figure \ref{fig:age_resonance}), and also found evidence that resonance fraction varies with planet radius in our `All Ages' sample (Figure \ref{fig:rp_resonance}),  we have not disentangled how resonance fraction depends on radius, mass, and age simultaneously. We argued that the younger planets approaching Jupiter in size were much lower than Jupiter in mass, as their gas envelopes may still be distended with the heat of formation (Section \ref{sec:sample}).  The masses of our sample are mostly not directly measured. Resonance fraction is expected to vary with gas content; we might expect more massive and gas-rich planets to occupy more resonances, both because they are formed in more massive gas disks which drive more extensive migration, and because more massive planets might better resist dynamical disruption. In this regard the ostensible disagreement between the low incidence of resonance that we measure for transiting giants, and the higher incidence of resonance reported for radial-velocity detected giants \citep{Wright2011}, should be examined.}


\end{itemize}

\section{Summary}
In this paper, we collected and analyzed young multi-planet systems with age constraints. Memberships in stellar clusters provided especially precise ages.
We identified near-resonant configurations between neighboring planets by evaluating how close their period ratios were to period commensurabilities (either 1st-order or 2nd-order). Planet pairs with sufficiently small deviations from perfect commensurability were labeled `resonant'. We studied the incidence of resonance across different ages, orbital architectures, and planetary radii (not always simultaneously). Our findings can be summarized as follows:

\begin{itemize}

    \item {Integrated over all ages, resonant pairs tend to congregate in planetary systems with high multiplicity and low mutual inclination. In systems where one 1st-order resonance is present, the probability of finding another 1st-order resonance is approximately doubled: from 15.6$\pm$1.1\% to 32.6$\pm3.4\%$.}


        \item{First-order resonances are more common than second-order resonances regardless of age (15.6$\pm$1.1\% vs. 5.8$\pm0.7\%$ when integrated over all ages). The 3:2 and 2:1 resonances are the most common 1st-order resonances. Resonances with period ratios closer to unity (4:3, 5:4, 6:5) are increasingly rarer. See Table~\ref{tab:mmr}.}

    \item{Young planetary systems ($<$100 Myrs old) seem to be dominated by resonant configurations, with 6/7 (86$\pm13$\%) having at least one resonant pair (including 1st and 2nd-order).  Among  neighboring pairs, 7/10 (70$\pm15$\%) are near 1st-order resonances. This prevalence of resonance among young planetary systems directly supports the `break-the-chains' models \citep{Izidoro,Izidoro2022,Goldberg2022}.}
    
    \item {The incidence of resonant configurations drops substantially for adolescent planets (between 100 Myrs and 1 Gyr in age), with 6/16 (38$\pm12$\%) for the fraction of resonant systems (including 1st and 2nd-order) and 7/29 (24.1$\pm7.9\%$) for the fraction of 1st-order resonant pairs. These rates of incidence are still lower for mature planets ($>$1-Gyr-old): 58/255 (22.7$\pm$2.6\%) and 59/402 (14.7$\pm$1.8\%), respectively. 
    }


      \item{The orbital periods of the innermost planets in resonant chains are predominantly $<$ 10 days, mirroring the rotation period distribution of young stars still hosting protoplanetary disks (see also \citealt{Lee_usp}). In contrast, the innermost planets of  generic multi-planet systems can have orbital periods longer than 100 days. These observations are broadly consistent with the idea that resonant chains were parked at the inner edges of their disks \citep[see e.g.][]{Wong2024}.}

 \item{The deviations from perfect period commensurability ($\Delta$) observed in the young planetary systems TOI-2076 (200 Myrs old; \citealt{Hedges}) and HD 109833 (10-16 Myrs old; \citealt{Wood}) are around 1\%. Resonant repulsion by equilibrium eccentricity tides would seem inadequate to explain such large deviations. Other processes such as planet-disk interactions \citep{Choksi2020}, stochastic forcing by disk turbulence \citep{Goldberg2023}, obliquity tides \citep{Millholland_obliquity}, and/or planetesimal scattering \citep{Chatterjee_2015, Wu2024} might be at play.}

 \item{ Super-Earths, especially those within the radius gap (\citealt{Fulton}; $1.5R_\oplus \leq R_p<1.9R_\oplus$), are less frequently observed to occupy a 1st-order resonance (21/187; 11.2$\pm2.3\%$) as compared to smaller Earth-sized planets ($R_p<1R_\oplus$; 16/79; 20.2$\pm4.5\%$) and larger sub-Neptunes ($1.9R_\oplus \leq R_p<2.5R_\oplus$; 33/212; 15.6$\pm2.7\%$). This lower incidence of resonance suggests a dynamically hot evolution for planets in the radius gap.}

\end{itemize}

\begin{acknowledgments}
We are grateful for insightful discussions with Dan Fabrycky, Man Hoi Lee, Eve Lee, Eric Agol, David Hernandez, Mutian Wang, Shuo Huang, Dan Tamayo, Kento Masuda, Ji-Wei Xie, Cristobal Petrovich, Yanqin Wu, and Wei Zhu.

\end{acknowledgments}

\bibliography{main}

\end{document}